\documentclass[article,twocolumn,prl,amssymb, superscriptaddress]{revtex4}

\usepackage{graphicx}
\usepackage{epsfig}
\usepackage{amsmath}
\usepackage{dcolumn}
\usepackage{amsmath}
\usepackage{latexsym}
\usepackage{bm}
\usepackage{color}
\usepackage[mediumspace,mediumqspace,Grey,squaren]{SIunits}

\begin{document}

\title{Specific Heat Signature of the Berezinskii-Kosterlitz-Thouless Transition in Ultrathin Superconducting Films}

\author{T.D. Nguyen}
\affiliation{Institut N\'eel, CNRS, BP 166, 38042 Grenoble Cedex 9, France.}
\author{A. Frydman}
\affiliation{The Department of Physics, Bar Ilan University, Ramat Gan 52900, Israel.}
\affiliation{Institut N\'eel, CNRS, BP 166, 38042 Grenoble Cedex 9, France.}
\author{O. Bourgeois}
\affiliation{Institut N\'eel, CNRS, BP 166, 38042 Grenoble Cedex 9, France.}
\affiliation{Univ. Grenoble Alpes, Inst NEEL, F-38042 Grenoble France}

\begin{abstract}
The Berezinskii-Kosterlitz-Thouless (BKT) transition is expected to have a clear signature on the specific heat. The singularity at the transition temperature $T_{BKT}$ is predicted to be immeasurable, and a broad non-universal peak is expected at $T>T_{BKT}$. Up to date this has not been observed in two-dimensional superconductors.  We use a unique highly sensitive technique to measure the specific heat of ultrathin Pb films. We find that thick films exhibit a specific heat jump at $T_C$ that is consistent with BCS theory. As the film thickness is reduced below the superconducting coherence length and the systems enters the 2D limit the specific heat reveals BKT-like behavior. We discuss these observations in the framework of the continuous BCS-BKT crossover as a function of film thickness.

\end{abstract}


\date{\today}

\maketitle

Within the 2D XY model, a second order phase transition cannot take place due to lack of long range phase coherence and the dominance of phase fluctuations (Goldstone modes).  Nevertheless, FBerezinskii and Kosterlitz-Thouless (BKT) \cite{berezinskii,KT} showed that a low-temperature quasi-ordered phase of bound vortex pairs exists leading to an infinite order phase-transition from bound vortex-antivortex pairs at low temperatures to unpaired vortices above the BKT critical temperature $T_{BKT}$. From the thermodynamic point of view BKT theory predicts that the specific heat, $c_p$,  is characterized by an immeasurable essential singularity at $T=T_{BKT}$ and a non-universal peak at $T > T_{BKT}$  associated with the liberation of entropy due to the unbounding of vortex-antivortex pairs \cite{chaikin}. Work on this transition led to the 2016 Nobel Prize in Physics being awarded to Kosterlitz and Thouless. 

\begin{figure}
   \begin{center}
  \includegraphics[width=0.5\textwidth]{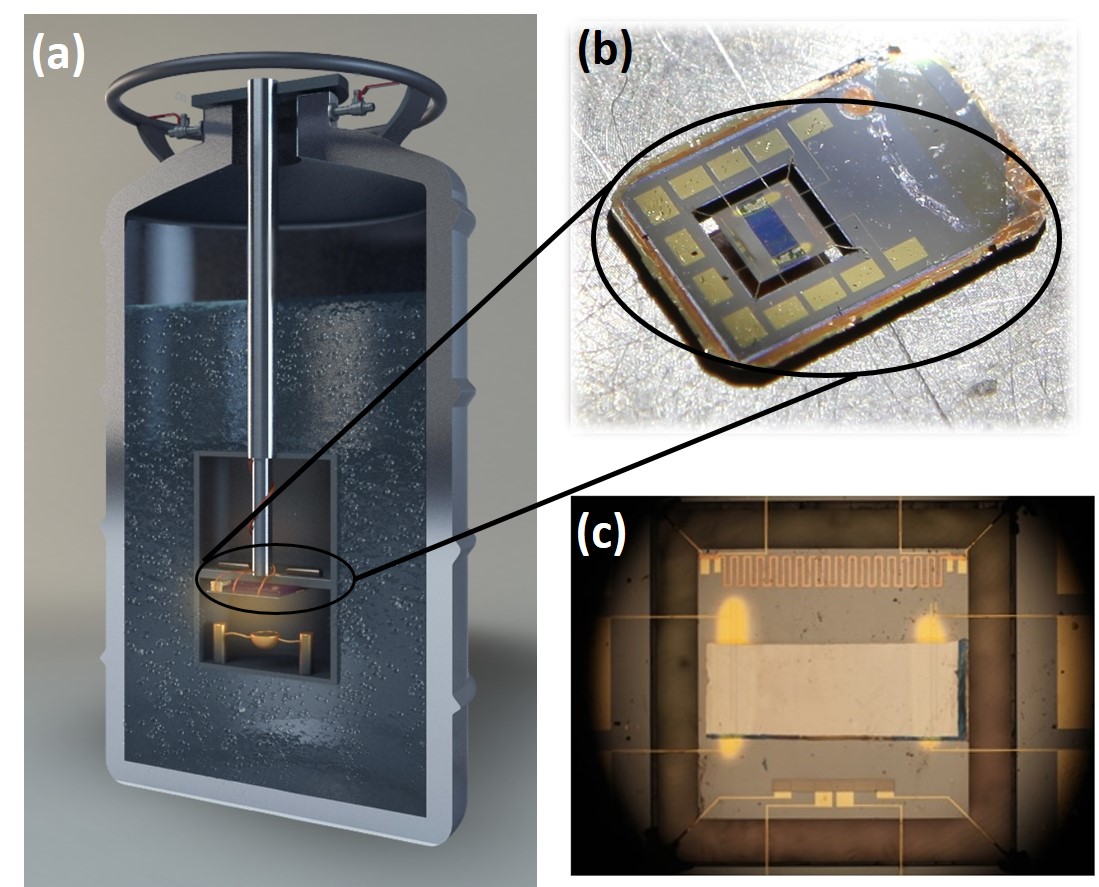}
    \caption{(a). The quench condensation set-up consists of evaporation baskets used for growing sequential continuous Pb layers, the substrate being held at cryogenic temperatures and in UHV conditions. (b) and (c), the suspended membrane acting as the thermal cell contains a copper meander, used as a heater, and a niobium nitride strip, used as a thermometer. These are lithographically fabricated close to the two edges of the active thermal sensor. The quench-condensed films are evaporated through a shadow mask which, together with the measurement leads, defines its geometry.} 
    \label{setup}
    \end{center}
    
 \end{figure}
 
A paradigmatic system in which the BKT transition may be expected is a 2D superconducting film. Evidences for the BKT physics have been reported in transport measurements via analysis of the $I$-$V$ characteristics or by studying the perpendicular magnetoresistance curves \cite{min,kadin,wolf,hebard}. However up to date there have been no experimental thermodynamic signatures of this transition especially concerning 2D superconducting films. This requires a highly sensitive thermal experiment which is able to resolve the specific heat of ultrathin films in the limit of 2D superconductivity \cite{bourgeois2005,poran,poran2}.
 
Here we report on specific heat, $c_p$, measurements performed on ultrathin superconducting films. We utilize a unique experimental setup based on a suspended silicone membrane substrate that enables to measure $c_p$ of Pb films with thicknesses ranging from 1.2~nm to 56~nm. We show that the thicker films can be well described by the BCS theory for strong coupled superconductors. In particular, they exhibit a specific heat jump at the critical temperature, $T_C$, characteristic of the second order phase transition. Much thinner films, on the other hand, do not posses a measurable jump at $T_C$ but are rather characterized by a broad $c_p$ peak at $T > T_C$ indicating the presence of an excess of entropy. These results are interpreted as thermodynamic signatures for a BCS-BKT crossover as a function of film thickness. 

The samples used in this work were sets of ultrathin Pb films having different thickness obtained by the quench condensation technique \cite{strongin,dynes1978,dynes1,Goldman89,olivier} i.e. sequential evaporations of ultrathin films on a cryogenically cooled substrate without thermal cycling to room temperature or exposing the film to atmosphere (see Fig.~\ref{setup}(a) and Supplemental Materials \cite{SM}). This allows in-situ sequential depositions under UHV conditions and simultaneous transport and thermal measurements on a single sample. Due to its unique advantages, this experimental method allows the study of the thermodynamic properties of the superconducting transition in ultra-thin layers as a function of thickness . 

The Pb thin films were evaporated layer by layer onto a calorimetric membrane sensor, after the deposition of an adhesion layer (0.5~nm of Sb) favoring the continuity of the superconducting films. The thermal sensor was composed of a thin silicon membrane with a thickness of about 5~$\mu$m, suspended by 12 arms for mechanical support as well as for electrical connections to the heater, thermometer and the evaporated sample on the membrane \cite{therm} (see Fig.~\ref{setup}(b) and (c)). Using this setup we were able to measure simultaneously the resistance per square $R_{sq}$ using four probe techniques and the heat capacity, $C_p$. 

The heat capacity was measured using an ac calorimetry technique, with sensitivity of a few tens of attoJoule per Kelvin \cite{bourgeois2005,poran,poran2}, of 22 sequential layers of Pb. The parameters of all layers are summarized in the table of the Supplementary Materials \cite{SM}, where more details of the experiment can be found.

\begin{figure}
\begin{center}
  \includegraphics[width=0.5\textwidth]{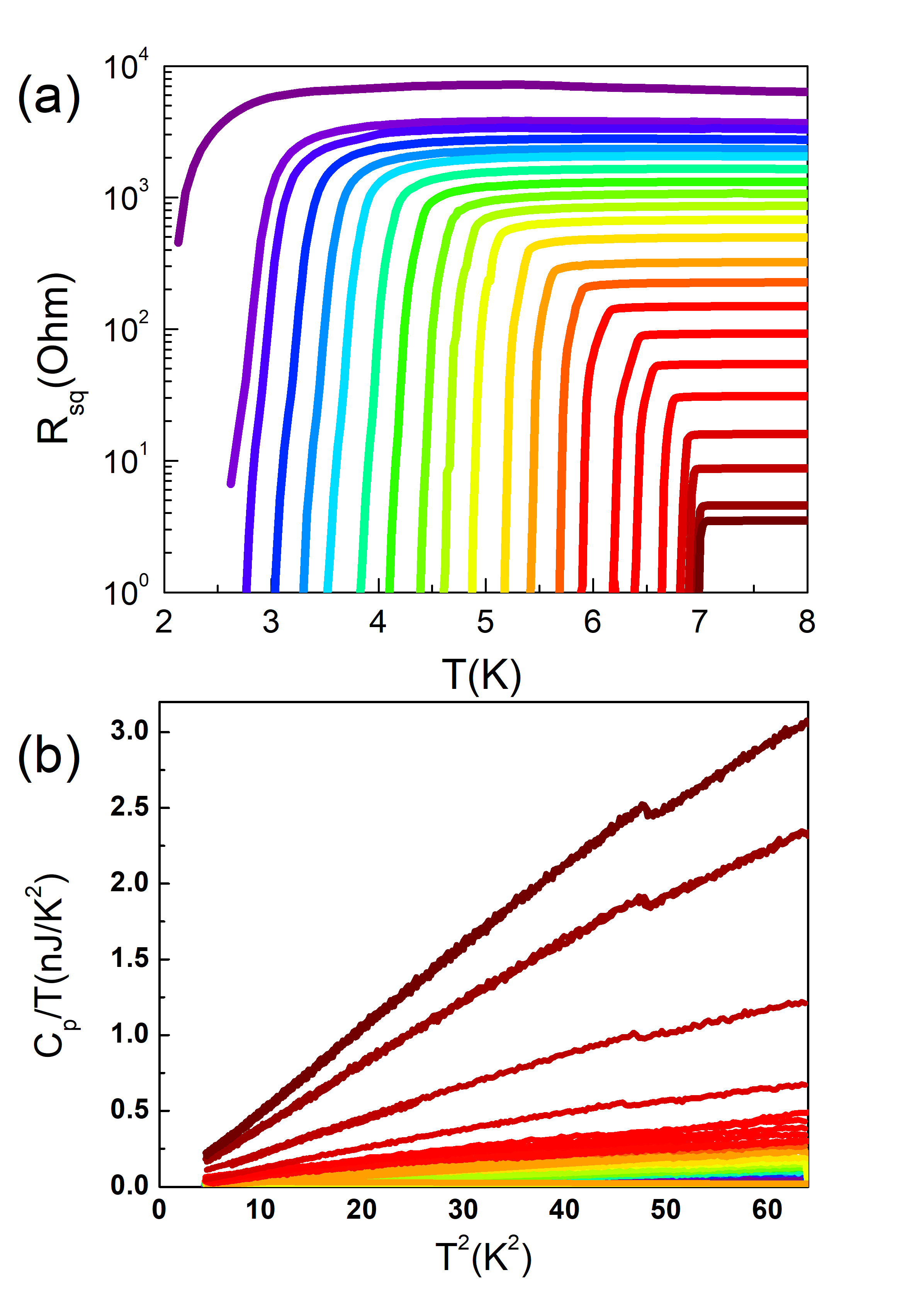}
     \caption{ (a) Resistance per square $R_{sq}$ versus temperature for the 22 sequential quench condensed lead films. Purple is for thin films and red-brown for thick films. This color code is maintained throughout the paper. (b) Heat capacity of the films (same color code) in form of $C_{p}/T$ versus $T^{2}$ highlighting the cubic behavior above 7.2~K.}
  
    \label{RT_CT}
    \end{center}
\end{figure}

Fig.~\ref{RT_CT}(a) shows resistance versus temperature curves of a set of quench condensed Pb films with thicknesses ranging between 1.2 and 56~nm. From these measurements, we extracted the critical temperature, $T_{C\it{res}}$, defined as the temperature at which resistance dropped to $10\%$ of its value at $T=10$~K.  $T_{C\it{res}}$ increased monotonically with increasing thickness of the lead layer, $t$. Our thinnest film ($t=1.22$~nm) exhibited  $T_{C\it{res}}=2.15$~K while films with $t \geq 12$~nm had critical temperatures close to that of bulk Pb $T_{C\it{bulk}}=7.2$~K. These values are in agreement with previous studies on ultrathin quench condensed Pb films \cite{Goldman89}. The heat capacity ($C_p$) measurements of the same films are shown in Fig.~\ref{RT_CT}(b), they are obtained after subtraction of the membrane heat capacity (Si, heater and thermometer, see Supplemental Materials \cite{SM}).

The heat capacity of a metallic sample is expected to follow the well known form:

\begin{equation}
\frac{C_n}{T} = \gamma + \beta T^2
\label{c_normal}
\end{equation}
where $\gamma$ and $\beta$ stand for the electron and phonon heat capacities coefficients respectively. For this reason the data is plotted as $C_{p}/T$ versus $T^{2}$ resulting in a linear normal-state curve above $T_C$. The heat capacity increased with film thickness and at high enough thickness, $t \geq 9$~nm (stage 18 and above), we observed a $C_p$ jump associated with the superconductor second order phase transition. The temperature position of the jump is consistent with the slight decrease of $T_{C\it{res}}$ with decreasing thickness in this regime (see Fig.~\ref{RT_CT}(a)). The amplitude of the jump, $\Delta C_{p}$, decreases with decreasing thickness until for $t \leq 9 $~nm  the jump becomes immeasurable, smaller than the noise. We note, however, that even for the thickest film (layer 22, $t=55.9$~nm) the ratio between the jump amplitude and the normal state heat capacity $C_p$ at $T_C$, $\Delta C_{p}/C_{n}(T_{c})=0.0445$ is much smaller than the expected BCS value of 1.4 obtained for bulk Pb for instance \cite{Shiffman2}. Like for Nb \cite{Brown} and Al \cite{Douglas}, this indicates that the heat capacity of amorphous Pb films is largely dominated by the phonon contribution. 

\begin{figure}
    \includegraphics[width=0.5\textwidth]{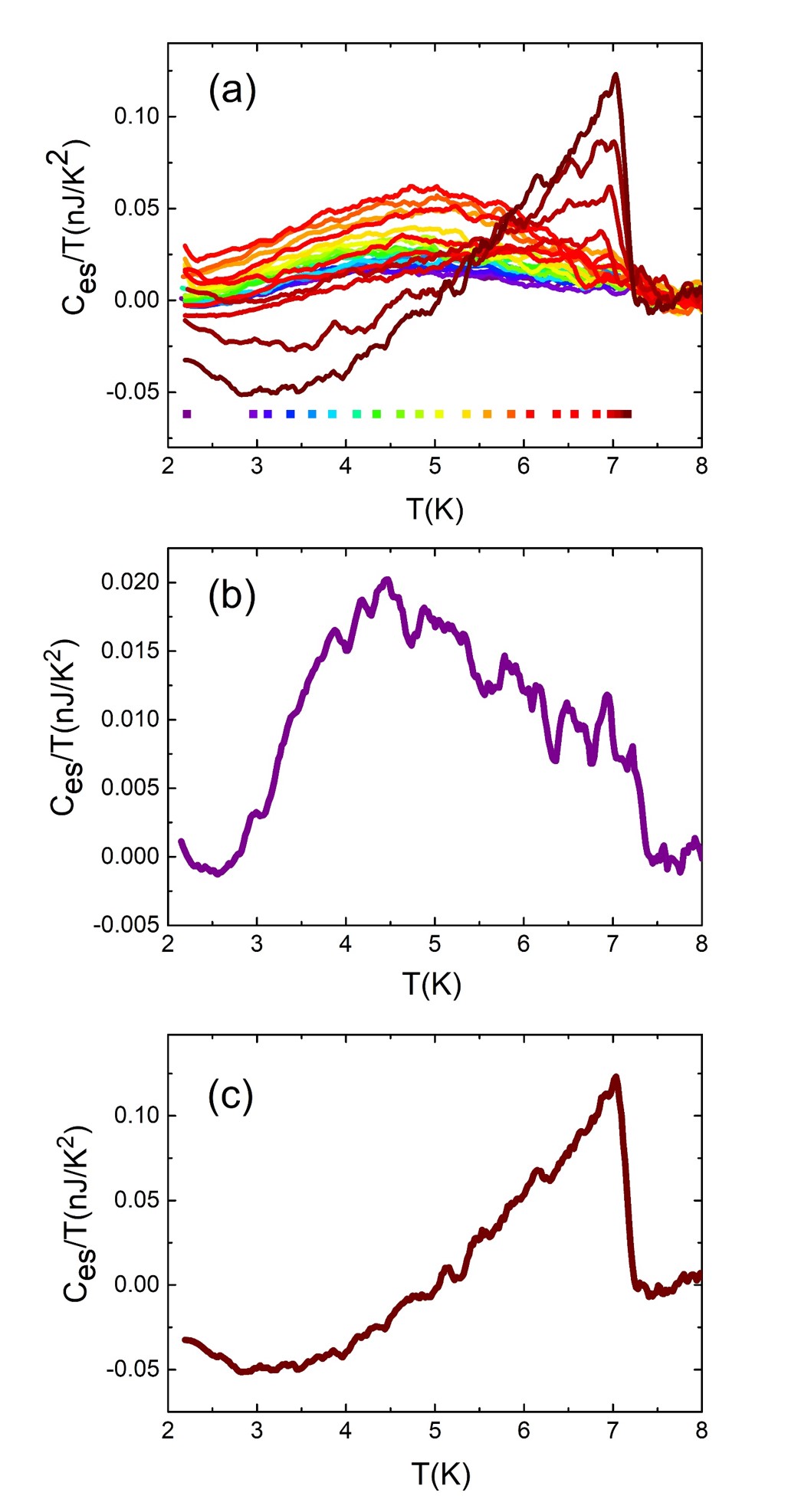}
    \vspace{-0.5cm}
    \caption{(a) Superconducting electronic heat capacity  $C_{es}$ of the films as extracted from the data presented in Fig.~\ref{RT_CT} along with an identical color code. The squares mark the $T_{C\it{res}}$ of each layer extracted from the RT curves of Fig.~\ref{RT_CT}(a). The curves for the 1.2~nm and 55~nm thick films are shown in (b) and (c) respectively.}  
    \label{BCS-BKT}
  \end{figure}

In order to focus only on the contribution to the heat capacity from electrons in the superconducting state $C_{es}$ we subtract the normal state $C_{p}$, extracted from the linear slope in $T^2$ above $T_{Cbulk}=7.2$~K, from each respective curve of Fig.~\ref{RT_CT}(b) thus obtaining  $C_{es}=C_p-C_n$. $C_{es}$ for the different layers are shown in Fig.~\ref{BCS-BKT}(a). $C_{es}$ versus $T$ for the thinnest and thickest films are shown in panels (b) and (c) of Fig.~\ref{BCS-BKT} respectively. The curves for the thickest films are consistent with results obtained on bulk Pb samples \cite{Shiffman2}. 

\begin{figure}
    \includegraphics[width=0.45\textwidth]{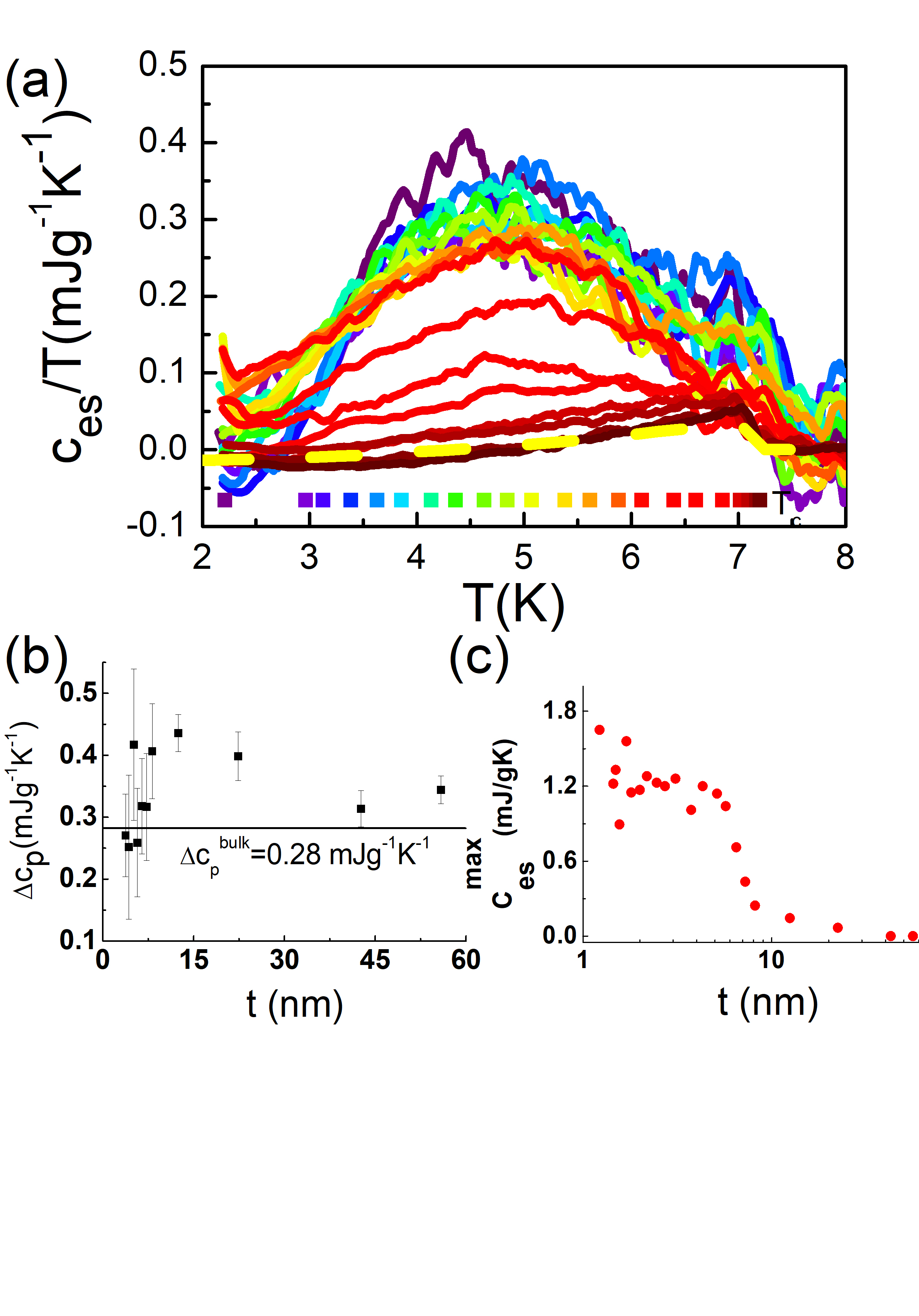}
    \vspace{-2cm}
    \caption{(a) Specific heat of electrons in the superconducting state $c_{es}$ versus $T$. The squares mark the $T_{C\it{res}}$ of each layer extracted from the RT curves of Fig.~\ref{RT_CT}(a). The yellow dashed line is a fit to BCS expectation (see Supplemental Materials \cite{SM}). (b) The amplitude of the specific heat anomaly jump at $T_C$, $\Delta c_p$, as a function of thickness. (c) The maximum value of the specific heat, normalized to the specific heat of layer 22, as a function of thickness. } 
    \vspace{-0.5cm}
    \label{3D}
  \end{figure}

For obtaining the specific heat, $c_p$, from the measured heat capacity $C_p$ can be achieved by dividing the curve of each layer in Fig.~\ref{BCS-BKT}(a) by its mass: $c_{p}^{i}=C_{p}^{i}/m^{i}$. $c_p$ versus $T$ curves for all layers are shown in Fig.~\ref{3D}.
It is illustrating that the specific heat jump amplitude for the thicker films is very close to that observed in bulk Pb samples, $\Delta c_p \sim 0.28$~mJ.g$^{-1}$.K$^{-1}$ as shown in the Fig.~\ref{3D}(b) (see also Supplemental Materials \cite{SM}). This is in stark contrast to results obtained on granular Pb films \cite{poran2} for which $\Delta c_p$ was found to be larger than the bulk value by up to a factor of eight. As the film is thinned, $\Delta c_{p}$ becomes immeasurable and an excess specific heat peak emerges with a temperature region that extends up to $T_{C\it{bulk}} = 7.2$~K. 
These results are consistent with a crossover from 3D BCS physics, characterized by $T_c=7.2$~K, and to 2D BKT physics with $T_{BKT} \approx 2$~K for the thinnest films while the intermediate layers show a mixture of both.

The representation shown in Fig.~\ref{3D} highlights the importance of the broad peaks which become more significant as the thickness of the layer is reduced. The magnitude of this peak increases sharply for $t \leq 10$~nm and saturates for $t \leq 5 $~nm as depicted in Fig.~\ref{3D}(c). This saturation of the specific heat peak amplitude for thicknesses below 5~nm is consistent with the superconducting film becoming 2D. The superconducting dirty limit coherence length, $\xi '$, of the amorphous Pb is given by $\sqrt{\xi_0 l}$ where $\xi_0$ is the clean limit coherence length (80~nm for Pb) and $l$ is the mean free path which for our samples is 0.3~nm \cite{olivierPRB}. This yields $\xi '=4.9$~nm. Hence, the excess specific heat bump reaches its full amplitude as the film thickness becomes comparable to the coherence length.

It should be noted that the amplitude of the specific heat peak at  $T > T_{BKT}$ is much larger than what could be naively expected if each vortex degrees of freedom contributes $2k_B$ to $C_p$ \cite{chaikin}. Assuming a single vortex per coherence length, $\xi '$, the measured peak amplitude is close to two orders of magnitude larger than the expected value.

A point to consider is related to the sample dimensionality. The electronic heat capacity includes contributions both from quasiparticles and from vortices. For quasiparticle the system has to be treated as 3D, and the specific heat should be obtained by dividing the heat capacity by the layer thickness (or by the mass). The  vortices, on the other hand, should organize in a 2D plane once $t<\xi '$, and hence the vortex contribution to  heat capacity is not expected to change with growing thickness. In this respect, it is interesting to compare Fig.~\ref{BCS-BKT}(a), which is representative of a 2D treatment, and  Fig.~\ref{3D}(a), which highlights 3D physics. One could expect that the heat capacity peak amplitude, $C_{es}^{max}$ in Fig.~\ref{BCS-BKT}(a) would not change with thickness for thin films. It should be noted, however, that ultrathin superconducting films have been shown to be characterized by "emergent electronic granularity" i.e. superconducting puddles embedded in an insulating matrix \cite{kowal,ghosal1,ghosal2,feigelman,Nandini}. These puddles may have different sizes and thus a spread of critical temperatures \cite{Biscaras}. This may be the reason why the specific heat in the thinnest films does not posses a jump at $T_C$ similar to the one observed in granular Pb samples where each grain is large enough to sustain bulk superconductivity \cite{poran2}. Thin enough layers may actually not achieve full coverage of the substrate, both morphologically and electronically. Increasing the thickness of film may increase the area of superconducting regions leading to increase the vortex contribution to heat capacity even in the 2D limit.   
 
In summary, we have successfully performed specific heat measurements on Pb films as thin as 1.2~nm having a mass as small as few tens of nanograms. 
We have shown that for the thicker films the specific heat jump is well described by the BCS model for strong coupled superconductivity. For the thinner films, a broad peak in $c_p$ is observed   without any measurable jump at the resistive critical temperature. 
These are quantitatively consistent with the BKT predictions in the limit of ultra-thin uniform superconducting films. Since the details of the specific heat versus temperature curves are predicted to be non-universal and are system dependent,  we are not able to compare our results to a quantitative model. Nevertheless, the amplitude of the specific heat signal is larger than expected from a naive estimation. 

We are grateful for useful discussions with M. Holzmann, N. Trivedi, and M. Randera and for the support from the technical pole of Institut NEEL especially from E. Andr\'e, T. Crozes, A. G\'erardin, G. Moiroux, and J.-L. Mocellin. We acknowledge support from the Laboratoire d’excellence LANEF in Grenoble (ANR-10-LABX-51-01). A.F. acknowledges support from the Israel US bi-national foundation grant no. 2014325. 
\vspace{1cm}

aviad.frydman@gmail.com
olivier.bourgeois@neel.cnrs.fr

\vspace{0.5cm}


\begin{thebibliography}{99}


\bibitem{berezinskii}
V.L. Berezinskii, Sov. Phys. JETP \textbf{34}, 610 (1972).

\bibitem{KT}
J.M. Kosterlitz, and D.J. Thouless, J. Phys. C \textbf{6} 1181 (1973).

\bibitem{chaikin}
P.M. Chaikin, and T.C. Lubensky. Principles of condensed matter physics, 550, Cambridge University Press (1995).

\bibitem{min} P. Minnhagen, Rev. Mod. Phys. \textbf{59}, 1001 (1987).

\bibitem{kadin} K. Epstein, A.M. Goldman, and A. M. Kadin, Phys. Rev. Lett. \textbf{47} 534 (1981).

\bibitem{wolf} S. A. Wolf, D. U. Gubser, W. W. Fuller, J. C. Garland, and R. S. Newrock, Phys. Rev. Lett. \textbf{47}, 1071 (1981).

\bibitem{hebard}  A. F. Hebard and A. T. Fiory, Phys. Rev. Lett. \textbf{50}, 1603 (1983).


\bibitem{bourgeois2005} O. Bourgeois, S.E. Skipetrov, F. Ong, and  J. Chaussy, Phys. Rev. Lett. \textbf{94}, 057007 (2005).

\bibitem{poran}  S. Poran, M. Molina-Ruiz, A. G\'erardin, A. Frydman, O. Bourgeois, Rev. Sci. Instrum. \textbf{85}, 053903 (2014).

\bibitem{poran2}  S. Poran, T. Nguyen-Duc, A. Auerbach, N. Dupuis, A. Frydman,  \&  O. Bourgeois,  Nat. Commun. \textbf{8}, 14464 (2017).

\bibitem{strongin} M. Strongin, R.S. Thompson, O.F. Kammerer, and J.E. Crow, Phys. Rev. B \textbf{1}, 1078 (1970).

\bibitem{dynes1978} R.C. Dynes, J.P. Garno, and J.M. Rowell, Phys. Rev. Lett. \textbf{40}, 479 (1978).

\bibitem{dynes1} R.C. Dynes, A.E. White, J.M. Graybeal, and J.P. Garno, 
Phys. Rev. Lett. \textbf{57}, 2195 (1986).


\bibitem{Goldman89} D.B. Haviland, Y. Liu, and A.M. Goldman,  Phys. Rev. Lett. \textbf{62}, 2180 (1989).



\bibitem{olivier} O. Bourgeois, A. Frydman, and R.C. Dynes, Phys. Rev. Lett. \textbf{88}, 186403 (2002).

\bibitem{SM}
See Supplemental Material at http://link.aps.org/  for details on the fabrication process, on the ac-calorimetry method used for the measurement of the heat capacity and on the Alpha-model used to do the BCS fit on the electronic specific heat in the superconductate.

\bibitem{therm} T. Nguyen, A. Tavakoli, S. Triqueneaux, R. Swami, A. Ruhtinas, J. Gradel, P. Garcia-Campos,  K. Hasselbach, A. Frydman, B. Piot, M. Gibert , E. Collin and O. Bourgeois, Journal of Low Temperature Physics, accepted (arXiv:1907.08443). 




A. Brown, M.W. Zemansky, and H.A. Boorse,  Phys. Rev. \textbf{92}, 52 (1953).


\bibitem{Douglas}
D.L. Martin, Proceedings of the Physical Society \textbf{78}, 1489 (1961).


\bibitem{olivierPRB}
O. Bourgeois, A. Frydman, and R.C. Dynes, Phys. Rev. B \textbf{68}, 092509 (2003).

\bibitem{kowal}	D. Kowal and Z. Ovadyahu, Solid State Comm. \textbf{90}, 783 (1994); \emph{ibid} Physica C \textbf{468} 322 (2008).

\bibitem{ghosal1}  A. Ghosal, M. Randeria, and N. Trivedi, Phys. Rev. Lett. \textbf{81}, 3940 (1998).

\bibitem{ghosal2}  A. Ghosal, M. Randeria, and N. Trivedi, Phys. Rev. B. \textbf{65}, 014501 (2001).

\bibitem{feigelman}  M.V. Feigel'man, L.B. Ioffe, V.E. Kravtsov, and E.A. Yuzbashyan, Phys. Rev. Lett. \textbf{98}, 027001 (2007).

\bibitem{Nandini} K. Bouadim, Y. Loh, M.  Randeria, and N. Trivedi, Nature Physics \textbf{7}, 884 (2011).


\bibitem{Biscaras}
J. Biscaras, N. Bergeal, S. Hurand, C. Feuillet-Palma, A. Rastogi, R.C. Budhani, M. Grilli,
S. Caprara, and J. Lesueur, Nat. Mat. \textbf{12}, 542 (2013).



\end{thebibliography}
\end{document}


\title{Supplemental Materials for Specific Heat Signature of the Berizinskii-Kosterlitz-Thouless Transition in Ultrathin Superconducting Films}

\author{T.D. Nguyen}
\affiliation{Institut N\'eel, CNRS, BP 166, 38042 Grenoble Cedex 9, France.}

\author{A. Frydman}
\affiliation{The Department of Physics, Bar Ilan University, Ramat Gan 52900, Israel.}
\affiliation{Institut N\'eel, CNRS, BP 166, 38042 Grenoble Cedex 9, France.}

\author{O. Bourgeois}
\affiliation{Institut N\'eel, CNRS, BP 166, 38042 Grenoble Cedex 9, France.}
\affiliation{Univ. Grenoble Alpes, Inst NEEL, F-38042 Grenoble France}

\pacs{}

\date{\today}

\maketitle
\maketitle

\textbf{S1. Sample fabrication and Experimental set-up} 

The quench condensation system consists of three thermal evaporators to deposit different materials on the Si membrane based calorimeter at cryogenic temperatures. For obtaining  continuous ultrathin films, a thin Sb adhesion layer of about 2.5~nm thick is evaporated onto the cryo-cooled substrate prior to the deposition of the first Pb layer. The first evaporated Pb layers is subnanometer thick, electrically insulating, having a heat capacity too low to be measured. We got a measurable $C_p$ signal for a layer of 1.2~nm for which the superconducting critical temperature is $T_{Cres}=2.12$~K. 

The layers are quench condensed on a uniquely designed 5~$\mu$m thick Si membrane based calorimeter suspended by 12 arms. The thermal sensor consists of a NbN thermometer 70~nm thick and a heater made of Cu (100~nm thick) installed on each side of the membrane to free space for the evaporated samples. All the microfabrication steps of the calorimeter are done using optical lithography. The electrical connections to all transducing elements on the membrane are obtained by a superconducting layer of (70~nm) NbTi/(20~nm) Au deposited on the suspending arms. In order to ensure a good electrical connection to very thin films (few angstroms), we evaporate (5~nm)WTi/(100~nm)Au on the contacts through a shadow mask to make the profile smooth.

The calorimeter is wire-bonded to a sample holder that is mounted on the quench-condensation system, a vacuum chamber immersed in liquid Helium and cooled to $T=2$~K. The sequential evaporations of Pb layers are carried out through a mechanical mask defining a window of 1.14~mm$\times$3.09~mm on the membrane, while temperature on the sample holder is regulated at 10~K during the material growth.

\textbf{S2. Heat capacity measurement technique} 

The heat capacity measurement has been performed using the ac-calorimetry technique, in which an ac current with frequency $f$ is applied to the heater, leading to the oscillation of the membrane temperature at the second harmonic $2f$ with amplitude of $\delta T_{\rm ac}$. Measuring the temperature oscillation enables us to extract the heat capacity using the equation:

\begin{equation}
C_{\rm p} = \frac{P_{\rm ac}}{4\pi f\delta T_{\rm ac}}
\end{equation}
with $P_{\rm ac}$ is the Joule heating power dissipated in the heater.

\begin{table*}
	\footnotesize
	\centering
	\begin{tabular}{|c|c|c|c|c|c|c|c|c|}
		\hline
		
		deposition $\sharp$ & mass ($\mu$g) & $C_{\rm p}$ (nJK$^{-1}$) & t(nm) & $\Delta C_{\rm p}$ (nJK$^{-1}$) & $\Delta c_{\rm p} $ (mJg$^{-1}$K$^{-1}$) &  R$_{\rm sq}$ (Ohm) & $T_{\rm c}$ (K)&  S$_{\rm 7K}$ (mJK$^{-1}$) \\
		\hline
		\hline
		1 & 0.04891 & 0.31433 & 1.22431 & NA & NA & 6436 & 2.15 & 1.08  \\
		\hline
		2 & 0.05789 & 0.5115 & 1.44923 & NA & NA &	3716 &	2.88& 0.815  \\
		\hline
		3 & 0.05953 & 0.5715 & 1.49018 & NA & NA & 3311 &  3.04 & 0.878  \\
		\hline
		4 &0.06242 & 0.58425 &	1.56269 &	NA &	NA &	2738 &	3.29 &	0.967  \\
		\hline
		5 & 0.0678&	0.633&	1.69733&	NA&	NA&	2316&	3.53&	1.07  \\
		\hline
		6 & 0.07198&	0.68175&	1.80188&	NA&	NA&	2046&	3.75&	0.914  \\
		\hline
		7 & 0.07988&	0.7575&	1.99973&	NA&	NA&	1636&	4.02&	1.03  \\
		\hline
		8 & 0.08701&	0.79275&	2.1783&	NA&	NA&	1304&	4.24&	0.997  \\
		\hline
		9 & 0.09815&	0.9&	2.45709&	NA&	NA&	1068&	4.50&	0.87  \\
		\hline
		10 & 0.10853&	1.035&	2.71694&	NA&	NA&	866&	4.71&	0.983  \\
		\hline
		11 &0.12348&	1.125&	3.09115&	NA&	NA&	680&	4.93&	0.826\\
		\hline
		12 & 0.14963&	1.3275 & 3.74573 & 0.04048 & 0.27055 & 478 &5.23& 0.785 \\
		\hline
		13 & 0.17192&	1.635&	4.30375&	0.04328& 0.25175& 323.2& 5.46& 0.967  \\
		\hline
		14 & 0.205&	1.8675&	5.13177&	0.0855&	0.41708&	225&	5.72&	0.851  \\
		\hline
		15 & 0.22791&	2.11592&	5.70546&	0.059&	0.25887& 148.4& 5.93& 0.876  \\
		\hline
		16 &0.25914&	2.4058&	6.48711&	0.08233&	0.3177&	91.1&	6.22&	0.606  \\
		\hline
		17 &0.2892&	2.6849&	7.23968&	0.09152&	0.31646&	44.6&	6.42&	0.35  \\
		\hline
		18 & 0.32558&	3.0227&	8.15055&	0.1324&	0.40665&	28.3&	6.66&	0.251  \\
		\hline
		19&	0.49785& 4.62197& 12.4629& 0.21696&	0.4358&	15.9& 6.82&	0.133 \\
		\hline
		20&	0.89413& 8.30106& 22.38336&	0.3562&	0.39838& 8.6& 6.90&	0.122 \\
		\hline
		21&	1.70387& 15.81861& 42.65404& 0.53438& 0.31363& 4.6& 6.97& 0.0365 \\
		\hline
		22&	2.23188& 20.72058&	55.87194& 0.7678& 0.34402& 3.5&	7.00& 0.0146 \\
		\hline
	\end{tabular}
	\caption{Experimental data extracted from the heat capacity measurements of the 22 evaporations (refereed to by sample $\sharp$). For each evaporation of Pb, we give the mass, the heat capacity $C_{\rm p}$ at 7.5~K, the thickness $t$, the heat capacity jump $\Delta C_{\rm p}$ at $T_{\rm c}$, the specific heat jump $\Delta c_{\rm p}$ at $T_{\rm c}$, the resistance per square R$_{\rm sq}$ and the $T_{\rm c}$ and the entropy of the superconducting electron $S_{7K}$ at 7~K extracted from the heat capacity measurements.}
	\label{table1}
\end{table*}

Prior to the first deposition, the heat capacity of the bare calorimeter (without sample) is measured in the temperature range from 2~K to 8~K. This was taken as a background for all consecutive layers. For each layer (including the Sb wetting layer) we simultaneously performed $R(T)$ and $C_p(T)$ measurements in the range 2~K to 8~K. For each stage, we extracted the specific heat by dividing the heat capacity by the layer mass:  $c_{p}^{i}=C_{Pb}^{i}/m^{i}$. The mass of the deposited Pb $m^{i}$ was determined by a quartz micro-balance integrated in the quench-condensation system and compared to the expected values from the superconducting transition temperature $T_{Cres}^{i}$ based on previous publication \cite{Goldman89}.  
All the experimental extracted from the $C_p$ measurement

\textbf{S3. Specific heat components}

The total specific heat has at least two components a phonon and an electron contributions. For this work, only matters the electronic contribution in the superconducting state to the specific heat. It is usually calculated using the following equation:
\begin{equation}
c_{es}=c_{s}-c_{n}
\end{equation}
where $c_{s}$ is the specific heat measured at zero field, which shows superconducting transition in the present case, $c_{s}=c_{p}^{i}$; $c_{n}$ is the specific heat in the normal state measured at magnetic field greater than the critical field. In our experiment, the critical field is expected to be much bigger than the limitation of our setup (the maximum available magnetic field is of 2~T) \cite{schiller}. In this case, we used an alternative strategy to estimate the electronic contribution. First, we fitted the specific heat of the normal state, above $T_{c}$ for stage 22 with a function: $c_{n}^{22}=\gamma*T+\beta*T^{3}+\zeta*T^{5}$, and then extrapolated to temperature below $T_{c}$ to find $c_{n}^{22}$ in the whole temperature range of the measurement from 2~K to 8~K. This $c_{n}^{22}$ was then used to estimate the electronic specific heat for all stages since the specific heat at normal state of all 22 stages are nearly overlapped. And so, the electronic specific heat for each stage is estimated by the following equation:

\begin{equation}
c_{es}^{i}=c_{p}^{i}-c_{n}^{22}
\end{equation}

\textbf{S4. Fitting the electronic specific heat with the $\alpha$-model}

It has been reported in number of works \cite{Shiffman1,Shiffman2} that bulk Pb is a strong-coupling superconductor, for which the BCS model does not fit the electronic specific heat very well. H. Padamsee and coworkers have developed an extended model based on BCS theory, the so called "$\alpha$-model" in 1973\cite{Shiffman1}. In this model, they introduced a free or adjustable parameter $\alpha\equiv\Delta(0)/k_{B}T_{c}$, which becomes a means of scaling the BCS gap:
\begin{equation}
\Delta(T)=(\alpha/\alpha_{BCS})\Delta_{BCS}(T)
\end{equation}
with $\alpha_{BCS}$ is the weak-coupling value of the gap ratio 1.764.
With this free parameter $\alpha$, the entropy of the superconducting electron becomes:
\begin{equation}
S_{es}(t)/\gamma T_{c}=-(3\alpha/\pi^{2})\int_{0}^{\infty}dx[f_{x}ln f_{x}+(1-f_{x}ln(1-f_{x})]
\end{equation}
where $f_{x}=[exp (\alpha t^{-1}(x^{2}+\delta^{2})^{1/2})+1]^{-1}$, $t=T/T_{c}$ and $\delta=\Delta(T)/\Delta(0)$ is the reduced gap. 

The specific heat of the superconducting electron is then calculated by the following equation:
\begin{equation}
C_{es}/\gamma T_{c}=t(d/dt)(S_{es}/\gamma T_{c}).
\end{equation}

\begin{figure}
	
	\includegraphics[width=0.5\textwidth]{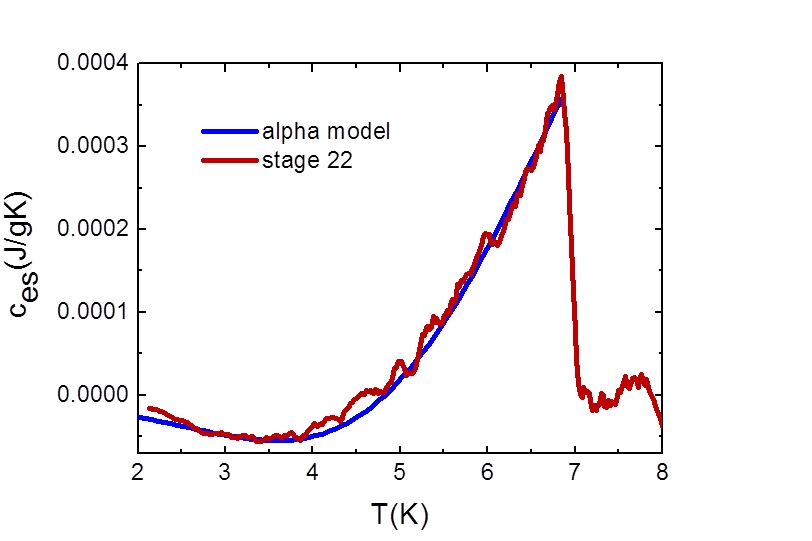}
	\caption{Fitting the electronic specific heat obtained from the last stage with $\alpha$-model.}
	\label{sfig:4}
\end{figure}

In order to fit the specific heat of the superconducting electron obtained at stage 22 ($c_{es}^{22}$), we firstly used Mathlab to calculate numerically the specific heat of the superconducting electrons based on the $\alpha$-model. This calculation gives us $c_{es}^{\alpha}$. Since the $c_{es}^{22}$ is obtained by removal from the $c_{p}^{22}$ the extrapolation of the normal state, which contains also the electronic contribution ($\gamma T$). Therefore to obtain the fit to our data, we have to subtract from the calculated specific heat a specific heat contribution coming from the normal electrons ($\gamma T$). It is also known that the $\gamma$ coefficient of strong-coupling superconductors like Pb is not a constant but temperature dependent \cite{Shiffman1,Shiffman2,Grimvall,Grimvall2}. Thus, we have fitted our data ($c_{es}^{22}$) with $c_{es}^{\alpha}-\gamma(T)T$. The fit is shown in Fig.S~\ref{sfig:4}. We found that the fit is in good agreement when we set $\alpha=2.7$, and the $\gamma$ is a temperature dependent function: $\gamma(T)=8*10^{-7}*T^{2}+10^{-5}$ (Jg$^{-1}$K$^{-2}$), in good agreement with what has been observed for bulk Pb in the past \cite{Shiffman1}.